\documentclass[journal]{IEEEtran}

\usepackage[OT1]{fontenc} 
\usepackage[numbers,sort&compress]{natbib}
\usepackage[cmex10]{amsmath}
\usepackage{amssymb}
\usepackage{bm}
\usepackage{braket}
\usepackage{graphicx}
\usepackage{color}
\usepackage{subfigure}
\usepackage{tabularx}
\usepackage{arydshln}
\usepackage{mathtools}
\usepackage{multirow}
\usepackage{algorithm,algorithmic}
\usepackage{url}
\usepackage{listings}
\usepackage{comment}
\usepackage{CJKutf8}

\def\CN{\mathcal{CN}}
\def\I{\mathbf{I}}

\def\H{\mathbf{H}}
\def\Hh{\hat{\mathbf{H}}}

\def\V{\mathbf{V}}

\def\X{\mathbf{X}}

\def\D{\mathbf{D}}

\newcommand{\minimize}{\mathop{\rm minimize}\limits}

\begin{document}
\title{Maximizing Spectrum Efficiency of Data-Carrying Reference Signals via Bayesian Optimization}

\author{
Taiki~Kato,~\IEEEmembership{Graduate Student Member,~IEEE},
Hiroki~Iimori,~\IEEEmembership{Member,~IEEE},\\
Chandan~Pradhan,~\IEEEmembership{Member,~IEEE},
Szabolcs~Malomsoky,
and Naoki~Ishikawa,~\IEEEmembership{Senior Member,~IEEE}.
}

\markboth{\today}
{Shell \MakeLowercase{\textit{et al.}}: Bare Demo of IEEEtran.cls for Journals}
\maketitle

\begin{abstract}
Data-carrying reference signals are a type of reference signal (RS) constructed on the Grassmann manifold, which allows for simultaneous data transmission and channel estimation to achieve boosted spectral efficiency at high signal-to-noise ratios (SNRs). However, they do not improve spectral efficiency at low to middle SNRs compared with conventional RSs. To address this problem, we propose a numerical optimization-based Grassmann constellation design on the Grassmann manifold that accounts for both data transmission and channel estimation. In our numerical optimization, we derive an upper bound on the normalized mean squared error (NMSE) of estimated channel matrices and a lower bound on the noncoherent average mutual information (AMI), and these bounds are optimized simultaneously by using a Bayesian optimization technique. The proposed objective function outperforms conventional design metrics in obtaining Pareto-optimal constellations for NMSE and AMI. The constellation obtained by our method achieves an NMSE comparable to conventional non-data-carrying RSs while enabling data transmission, resulting in superior AMI performance and improved spectral efficiency even at middle SNRs.
\end{abstract}

\begin{IEEEkeywords}
Channel estimation, Grassmann manifold, 
manifold optimization, noncoherent multiple-input multiple-output (MIMO) communication.
\end{IEEEkeywords}

\IEEEpeerreviewmaketitle

\section{Introduction}
Global mobile data traffic is on a continuous upward trend and is expected to grow approximately threefold between 2024 and 2029 \cite{2024ericsson}.
Within this growth, the share of 5G new radio (NR) is expected to increase, accounting for $75\%$ of the total by 2029 \cite{2024ericsson}.
In 5G NR, five types of reference signals (RSs) are implemented, enabling accurate channel estimation and tracking. 
Among them, the demodulation RSs can account for up to approximately $30\%$ of the overall transmission signals \cite{5gnrch}.
For example, in a high-mobility scenario such as a bullet train with a travel speed exceeding $300\ \mathrm{km/h}$, the channel changes extremely rapidly over time \cite{ai2014challenges}. 
In such scenarios, frequent channel estimation is required, which increases the proportion of RSs and could potentially lower spectral efficiency.

Reducing RSs is one of the main approaches to improving spectral efficiency effectively. 
Known methods include a semi-blind method that reduces RSs through iterative channel estimation \cite{medles2001semiblind}, a blind method that estimates the channel without RSs \cite{shahbazpanahi2005closed}, and a superimposed pilot method that embeds RSs into data symbols \cite{hoeher1999channel}. 
All these methods detect data on the basis of the estimated channel, and such communication is referred to as coherent communication.
When the channel changes rapidly, improving channel estimation accuracy is crucial for enhancing spectral efficiency \cite{medard2000effect}.

Another approach to improving spectral efficiency is noncoherent communication, which does not require channel estimation. 
Well-known enablers include unitary space-time modulation \cite{hochwald2000unitary,hochwald2000differential}, differential coding \cite{weber1978differential,tarokh2000dostbc,ishikawa2018differential}, and methods based on the Grassmann manifold \cite{zheng2002communication}.
A key game-changer here uses reference signals derived from unitary space-time codewords \cite{yu2007informationbearing}, making it possible to perform channel estimation and data transmission simultaneously.
Building on this concept, RSs on the Grassmann manifold have been proposed \cite{endo2024boosting}.
These data-carrying RSs (DC-RSs) can potentially replace the conventional RSs in classical training methods and boost spectral efficiency at high signal-to-noise ratios (SNRs).
However, at low to middle SNRs, conventional RSs can still offer higher spectral efficiency as DC-RSs become less advantageous when the proportion of coherently detected data symbols increases.
This improvement can also be achieved even in high-mobility scenarios \cite{kato2024performance}.

In noncoherent communication using the Grassmann manifold, information is assigned to its subspaces.
Because multiplying a space-time codeword on the manifold by a channel matrix does not change its subspace, the point on the manifold remains the same, allowing for simultaneous data and channel estimation.
Designing a constellation on the Grassmann manifold is known as the Grassmann subspace packing problem \cite{conway1996packing}.
Although the principle of maximizing the minimum distance between subspaces appears straightforward, this problem is highly challenging. Over the years, numerous Grassmann constellation design methods have been introduced \cite{kammoun2003new, kammoun2007nonCoherent, gohary2009noncoherent, ashikhmin2010grassmannian, attiah2016systematic, elmossallamy2019noncoherent, ngo2020cubeSplit, cuevas2023constellations, cuevas2023union}.

Generally, construction methods fall into two categories: algebraic construction and numerical optimization.
A representative algebraic approach is exponential map (Exp-Map) \cite{kammoun2003new, kammoun2007nonCoherent}.
More recently, the remarkable Cube-Split \cite{ngo2020cubeSplit} and Grass-Lattice \cite{cuevas2023constellations} constellations have been proposed for single-input multiple-output (SIMO) scenarios, achieving higher minimum distances.
These algebraic methods are appealing because they enable the design of low-complexity detectors.

By contrast, numerical optimization-based approaches include methods that maximize the minimum distance between codewords \cite{gohary2009noncoherent, elmossallamy2019noncoherent} or minimize an upper bound on the pairwise error probability (PEP) \cite{cuevas2023union}.
Another notable technique integrates numerical optimization with algebraic construction \cite{attiah2016systematic}, combining the strengths of both methods.
Unlike purely algebraic approaches, numerical optimization makes it possible to obtain constellations tailored to any desired objective function, allowing flexible setting of parameters such as the numbers of transmission bits and time slots, and facilitating straightforward extension to multiple-input multiple-output (MIMO) scenarios.
Ideally, directly optimizing performance metrics such as channel estimation accuracy and average mutual information (AMI) would yield a desirable constellation.
However, Monte Carlo simulations are required for these calculations, and it is impractical to use them as the objective in numerical optimization.
Consequently, appropriate closed-form objective functions are required for practical numerical optimization.

Against this background, we propose a numerical optimization approach to designing the Grassmann constellation that maximizes the total AMI in communication using DC-RSs, and its maximization leads to maximizing spectral efficiency. Here, the total AMI is defined as the sum of (1) the AMI obtained from non-coherent detection when DC-RSs are transmitted and (2) the AMI obtained from coherent detection, using the channel estimated from the DC-RSs, when data symbols are transmitted subsequently. The contributions of this paper are as follows.
\begin{enumerate}
    \item We clarify that the total AMI of communication with DC-RSs depends only on three factors: the channel estimation accuracy of DC-RSs, the noncoherent AMI, and SNR. 
    From this relationship, it is shown that by simultaneously optimizing the channel estimation accuracy and the noncoherent AMI of DC-RSs, we can maximize the total AMI of communication, which is valid regardless of the SNR.
    \item We propose an objective function that enables the simultaneous optimization of the channel estimation accuracy and the noncoherent AMI from DC-RSs, both of which have a fundamental trade-off.
    In this process, a Bayesian optimization technique is used to efficiently search for good parameters.
    We demonstrate that the standard design metric is not suitable for optimizing the total AMI of communication.
    To address this problem, we derive an upper bound for the normalized mean squared error (NMSE) of estimated channel matrices, and a lower bound for the noncoherent AMI, proposing a new objective function that optimizes these bounds.
    \item We demonstrate that constellations on the Pareto front for NMSE and noncoherent AMI can be obtained efficiently by using the proposed objective function, and the constellation obtained through the proposed optimization method achieves better performance in terms of channel estimation accuracy and total AMI than all the conventional Grassmann constellations.
    Moreover, the results show that the proposed method achieves performance equal to or better than that of the classical training method using conventional RSs across all SNRs, indicating the effectiveness of using DC-RSs for all SNRs.
\end{enumerate}

The remainder of this paper is organized as follows.
In section $\mathrm{I\hspace{-1.2pt}I}$, we review conventional channel estimation methods and evaluation metrics of coherent communication.
In section $\mathrm{I\hspace{-1.2pt}I\hspace{-1.2pt}I}$, we introduce conventional construction methods for Grassmann constellations, evaluation metrics of noncoherent communication, and conventional DC-RSs.
In section $\mathrm{I}\hspace{-1pt}\mathrm{V}$, we present the proposed objective functions and optimization method.
Then, in section $\mathrm{V}$, we evaluate the performance of the constellation obtained using the proposed optimization method.
Finally, in section $\mathrm{V}\hspace{-1pt}\mathrm{I}$, we conclude this paper.

\section{System Model}
\label{sec:sys}
In this section, we describe the channel model assumed in this paper as well as a conventional channel estimation method.
NMSE is used as a metric of channel estimation accuracy, and AMI as a metric of maximum information rate that can be conveyed successfully in channel-coded scenarios.
We also review the NMSE of estimated channel matrices and the coherent AMI assuming channel estimation errors.

\subsection{Channel Model}
We consider a MIMO system with $M$ transmit antennas and $N$ receive antennas.
The channel model is quasi-static Rayleigh fading, where the channel matrix $\mathbf{H}\in\mathbb{C}^{M\times N}$ remains constant for $T$ time slots and each element of $\mathbf{H}$ follows $\CN(0,1)$. 
A space-time block code (STBC), $\mathbf{S}\in\mathbb{C}^{T\times M}$, satisfies the power constraint $\mathrm{E}[\|\mathbf{S}\|^{2}_{\mathrm{F}}]=T$. 
The received symbol $\mathbf{Y}\in\mathbb{C}^{T\times N}$ is given by
\begin{align}
    \mathbf{Y}=\mathbf{S}\mathbf{H}+\sigma_{\mathrm{v}}\mathbf{V},
    \label{systemmodel}
\end{align}
where each element of $\mathbf{V}\in\mathbb{C}^{T\times N}$ follows $\CN(0,1)$, and the SNR is defined as $\mathrm{SNR}=10\cdot\mathrm{log}_{10}{(1/\sigma_{\mathrm{v}}^{2})}\ [\mathrm{dB}]$.

\subsection{Conventional Channel Estimation and Data Detection}
In coherent scenarios, the channel matrix is estimated by using known RS, and data is detected on the basis of the estimated channel.
Owing to its low complexity nature, the training method is used in typical communication standards.
Let $T=T_{\mathrm{p}}$ denote the number of time slots for the RS, and assume that the RS $\mathbf{P}\in\mathbb{C}^{T_{\mathrm{p}}\times M}$ is known at both the receiver and the transmitter.
At the receiver, the channel is estimated using the known RS $\mathbf{P}$ and the received signal $\mathbf{Y}$. 
In the zero-forcing (ZF) case, the estimated channel $\hat{\mathbf{H}}$ is obtained using the pseudo-inverse matrix $\mathbf{P}^{+}=(\mathbf{P}^{\mathrm{H}}\mathbf{P})^{-1}\mathbf{P}^{\mathrm{H}}$, whereas in the case of minimum mean square error (MMSE) estimation, it is obtained using a weight matrix, i.e., \cite{endo2024boosting}
\begin{align}
    \hat{\mathbf{H}}=\begin{cases}
    \mathbf{P}^{+}\mathbf{Y}& \textrm{(ZF)} \\
    (\mathbf{P}^{\mathrm{H}} \mathbf{P} + \sigma^2_{\mathrm{v}}\mathrm{I}_M)^{-1} \mathbf{P}^{\mathrm{H}}\mathbf{Y}& \textrm{(MMSE)} \\
    \end{cases}.
    \label{channel_est_DMRS}
\end{align}

The receiver performs maximum likelihood detection using $\hat{\mathbf{H}}$, and the estimated symbol $\hat{\mathbf{S}}$ is given by
\begin{align}
    \hat{\mathbf{S}}=\underset{\mathbf{S}\in\mathcal{S}}{\mathrm{argmin}}\ \|\mathbf{Y}-\mathbf{S}\hat{\mathbf{H}}\|^2_\mathrm{F},
\end{align}
where $\mathcal{S}$ represents a codebook of space-time codewords.
Although maximum likelihood detection enables optimal symbol estimation, the computational complexity increases exponentially with the number of transmission bits.
In practice, MMSE equalization is a standard method to detect symbols independently because of its low computational complexity.

\subsection{Channel Estimation Accuracy}
Channel estimation accuracy is evaluated from the difference between the actual channel and estimated channel matrices, with NMSE used as the metric.
To prevent the increase in NMSE at low SNRs, the estimated channel $\hat{\mathbf{H}}$ is normalized as $\bar{\mathbf{H}}=\hat{\mathbf{H}}/\alpha_{\mathrm{c}}$, where $\alpha_{\mathrm{c}}=\sqrt{\mathrm{E}[\|\hat{\mathbf{H}}\|_\mathrm{F}^2]/ (N\cdot M)}$.
Using this normalized estimated channel $\bar{\mathbf{H}}$, we express NMSE as
\begin{align}
    \mathrm{NMSE}=10\cdot\mathrm{log}_{10}(\sigma_{\mathrm{e}}^2)~\mathrm{[dB]},\label{NMSE_sigma}
\end{align}
and
\begin{align}
    \sigma_{\mathrm{e}}^2=\frac{\mathrm{E}[\|\bar{\mathbf{H}}-\mathbf{H}\|_\mathrm{F}^2]}{\mathrm{E}[\|\mathbf{H}\|_\mathrm{F}^2]}.\label{NMSE}
\end{align}
Here, we model the channel estimation error by using the Gauss-Markov uncertainty model from \cite{nosrat-makouei2011mIMO}.
The estimated channel $\bar{\mathbf{H}}'$, which includes an error, is given by \cite{endo2024boosting}
\begin{align}
    \bar{\mathbf{H}}'=\sqrt{1-\beta_{\mathrm{c}}^2}\mathbf{H}+\beta_{\mathrm{c}}\mathbf{E},\label{CSI_error}
\end{align}
where $\mathbf{E}\in \mathbb{C}^{M \times N}$ is an error matrix with each element following $\CN(0,1)$.
The parameter $0\leq\beta_{\mathrm{c}}\leq1$ represents the uncertainty of $\bar{\mathbf{H}}'$, where $\beta_{\mathrm{c}}=0$ means that the estimated channel and the actual channel are completely identical. 
From \eqref{NMSE} and \eqref{CSI_error}, the relationship between NMSE and $\beta_{\mathrm{c}}$ can be respectively written as
\begin{align}
    \sigma_{\mathrm{e}}^2= \frac{\mathrm{E}[\|\bar{\mathbf{H}}'-\mathbf{H}\|_{\mathrm{F}}^2]}{\mathrm{E}[\|\mathbf{H}\|_{\mathrm{F}}^2]}=2\cdot(1-\sqrt{1-\beta_{\mathrm{c}}^2}),\label{NMSE_beta}
\end{align}
and
\begin{align}
    \beta_{\mathrm{c}}=\sqrt{1-\left(1-\frac{\sigma_{\mathrm{e}}^2}{2}\right)^2}.
\end{align}

\subsection{Coherent AMI with Channel Estimation Errors}
Let us consider a codebook $\mathcal{S}$ with the cardinality $|\mathcal{S}|=2^{B_{\mathrm{d}}}$, and each codeword $\mathbf{S}_i\in \mathbb{C}^{T_{\mathrm{d}} \times M}$ of $\mathcal{S}$ conveys $B_{\mathrm{d}}$ bits.
According to the system model \eqref{systemmodel}, the received symbol is represented as $\mathbf{Y}_i=\mathbf{S}_i\mathbf{H}+\sigma_{\mathrm{v}}\mathbf{V}$.
Here, the coherent AMI with channel estimation errors is given by \cite{ng2006mimo,endo2024boosting}
\begin{align}
R_{\mathrm{d}}=\frac{B_{\mathrm{d}}}{T_{\mathrm{d}}}-\frac{1}{T_{\mathrm{d}}|\mathcal{S}|}\mathrm{E}_{\bar{\mathbf{H}}',\mathbf{V}}\left[\sum_{i=1}^{|\mathcal{S}|}\log_2\left(\frac{\sum_{j=1}^{|\mathcal{S}|}p(\mathbf{Y}_i|\mathbf{S}_j,\bar{\mathbf{H}}')}{p(\mathbf{Y}_i|\mathbf{S}_i,\bar{\mathbf{H}}')}\right)\right],\label{AMI_d}
\end{align}
where \cite{endo2024boosting}
\begin{align}
    p(\mathbf{Y}|\mathbf{S},\bar{\mathbf{H}}')=\frac{1}{(\pi(\sigma_{\mathrm{v}}^2+\sigma_{\mathrm{e}}^2))^{NT_{\mathrm{d}}}}\exp\left(-\frac{\|\mathbf{Y}-\mathbf{S}\bar{\mathbf{H}}'\|_{\mathrm{F}}^2}{\sigma_{\mathrm{v}}^2+\sigma_{\mathrm{e}}^2}\right),\label{AMI_d_p}
\end{align}
and \cite{endo2024boosting}
\begin{align}
\frac{p(\mathbf{Y}_i|\mathbf{S}_j,\bar{\mathbf{H}}')}{p(\mathbf{Y}_i|\mathbf{S}_i,\bar{\mathbf{H}}')}=\exp \left(\frac{- z_{i,j}+z_{i,i} }{\sigma_{\mathrm{v}}^2+\sigma_{\mathrm{e}}^2}\right).
\label{AMI_d_pp}
\end{align}
Here, $z_{i,j}$ can be calculated as \cite{endo2024boosting}
\begin{align}
z_{i,j}\simeq\|(\mathbf{S}_i-\sqrt{1-\beta_{\mathrm{c}}^2}\mathbf{S}_j)\mathbf{H}+\sqrt{\sigma_{\mathrm{v}}^2+\beta_{\mathrm{c}}^2}\V\|_{\mathrm{F}}^2.\label{z_ij}
\end{align}

\section{Noncoherent Communication Based on Grassmann Constellation}
In this section, we describe communication based on Grassmann constellation.
First, we define the Grassmann manifold and introduce representative Grassmann constellations.
Next, we explain noncoherent detection using the representative Grassmann constellation and review the method for calculating AMI in noncoherent communication. 
Finally, we describe the conventional DC-RSs, which use noncoherent communication for channel estimation, and methods to improve its channel estimation accuracy.

\subsection{Grassmann Manifold}
For nonnegative integers $k$ and $n\geq k$, the Grassmann manifold $\mathcal{G}(n,k)$ represents the set of all $k$-dimensional subspaces in an $n$-dimensional space. 
In this paper, we define the Grassmann manifold as a quotient space of the Stiefel manifold by the $k \times k$ unitary group:
\begin{align}
\mathcal{U}(k) = \{\mathbf{U} \in \mathbb{C}^{k \times k} ~ | ~ \mathbf{U}^{\mathrm{H}}\mathbf{U} = \mathbf{I}_k\}.
\end{align}
The Stiefel manifold $\mathcal{S}(n,k)$ is the set of all $n \times k$ orthonormal basis matrices and is represented by 
\begin{align}
\mathcal{S}(n,k) = \{\mathbf{S} \in \mathbb{C}^{n\times k} ~ | ~ \mathbf{S}^{\mathrm{H}}\mathbf{S} = \mathbf{I}_k \}.
\end{align}
Here, let the equivalence relation $\sim$ signify that subspaces are equal. 
In the Stiefel manifold, the set of matrices that are equivalent under this relation is denoted by
\begin{align}
[\mathbf{S}] = \{\mathbf{S}_1 \in \mathcal{S}(T,M) ~ | ~ \mathbf{S}_1 \sim \mathbf{S}\}.
\label{eq:equivalentGr}
\end{align}
Operations on the Stiefel manifold that leave the subspace unchanged are equivalent to multiplication by a unitary matrix from the right.
Thus, when represents the subspace of the Grassmann manifold with an orthonormal basis, the Grassmann manifold can be defined as
\begin{align}
\mathcal{G}(n,k) = \mathcal{S}(n,k) ~ / ~ \mathcal{U}(k) = \{[\mathbf{S}] ~ | ~ \mathbf{S} \in \mathcal{S}(n,k)\}.
\label{eq:GTM}
\end{align}
When using points on the Grassmann manifold as codewords, we set $n=T$ and $k=M$, i.e., each subspace representing a codeword. 
Later, a Grassmann codebook is denoted by $\mathcal{X}=\{\mathbf{X}_i\in\mathcal{G}(T,M)\ |\ i=1,\cdots,2^{B_{\mathrm{g}}}\}$.

\subsection{Conventional Grassmann Constellation}
We introduce conventional construction methods for the Grassmann constellation, including Exp-Map, which is based on exponential mapping, Grass-Lattice, an algebraic construction for SIMO, and a method for maximizing the minimum distance through numerical optimization.

\subsubsection{Exp-Map \cite{kammoun2003new,kammoun2007nonCoherent}}
Using the matrix exponential, we can represent any point on the Grassmann manifold as
\begin{align}
    \mathbf{X}=\left[\exp{
    \begin{pmatrix}
    \mathbf{0} & \mathbf{C}_{i} \\
    - \mathbf{C}_{i}^{\mathrm{H}} & \mathbf{0}
    \end{pmatrix}
    } \right]\mathbf{I}_{T,M},\label{expmap}
\end{align}
where $\mathbf{I}_{T,M}=[\mathbf{I}_{M\times M}\ \mathbf{0}_{M\times(T-M)}]^\mathrm{T}$ and $\mathbf{C}_{i}\in\mathbb{C}^{M\times(T-M)}$ is an arbitrary matrix.
One method for constructing the constellation is to map QAM symbols to $\mathbf{C}_{i}$, for example. 
The constellation generated by exponential mapping is characterized by its high channel estimation accuracy \cite{endo2024boosting}.

\subsubsection{Grass-Lattice \cite{cuevas2023constellations}}
The construction of Grass-Lattice in SIMO is based on a mapping $\mathcal{M}$ from a unit hypercube to the Grassmann constellation.
The unit hypercube is the Cartesian product of the interval $(0,1)$ taken $2(T-1)$ times, and the interval $(0,1)$ is evenly divided into $2^{B_{\mathrm{l}}}$ equal segments.
Thus, the bit length of Grass-Lattice is $B_{\mathrm{g}}=2(T-1)B_{\mathrm{l}}$.
The mapping $\mathcal{M}$ consists of three mappings, written as
\begin{align}
    \mathcal{M}=\mathcal{M}_3\circ\mathcal{M}_2\circ\mathcal{M}_1.
\end{align}
First, in $\mathcal{M}_1$, points on the unit hypercube are mapped to the complex Gaussian distribution $\CN(0,1)$. 
Next, $\mathcal{M}_2$ maps points following a complex Gaussian distribution to the unit sphere. 
Finally, the points on the unit sphere are mapped onto the Grassmann manifold through $\mathcal{M}_3$. 
For the open interval $(0,1)$, it is necessary to define a lower bound $\alpha_{\mathrm{l}}$ and an upper bound $1-\alpha_{\mathrm{l}}$.
It can be optimized for performance, and \cite{cuevas2023constellations} provides the optimal $\alpha_{\mathrm{l}}$ for specific parameters.

\subsubsection{Manifold Optimization for Maximizing Minimum Chordal Distance (Mopt-MCD) \cite{gohary2009noncoherent,ngo2020cubeSplit}}
Maximizing the minimum distance between symbols can improve the symbol error rate (SER) in coherent codes.
By appropriately defining the distance, we can also improve the SER of noncoherent codes \cite{gohary2009noncoherent}.
Here, we consider chordal distance as a distance metric for the Grassmann constellation \cite{ngo2020cubeSplit}. 
The chordal distance $d_\mathrm{c}(\mathbf{X}_i,\mathbf{X}_j)$ between two codewords is given by \cite{ngo2020cubeSplit}
\begin{align}
    d_\mathrm{c}(\mathbf{X}_i,\mathbf{X}_j) := \sqrt{M-\|\mathbf{\X}_{i}^{\mathrm{H}}\X_j\|_{\mathrm{F}}^2}.\label{chordal_distance}
\end{align}
The constellation obtained by maximizing the minimum chordal distance (MCD) demonstrates the best performance in terms of noncoherent AMI \cite{endo2024boosting}.
From \eqref{chordal_distance}, the MCD can be maximized by minimizing the maximum value of $\|\mathbf{\X}_{i}^{\mathrm{H}}\X_j\|_{\mathrm{F}}^2$. 
The max function can be approximated by using the log-sum-exp function to make it a differentiable function, enabling continuous optimization on the manifold \cite{boumal2014manopt}.
Thus, the problem of maximizing MCD can be represented as minimizing the metric $f_{\mathrm{MCD}}$, written as \cite{endo2024boosting}
\begin{align}
    f_{\mathrm{MCD}}:=\mathrm{log}\sum^{|\mathcal{X}|-1}_{i=1}\sum^{|\mathcal{X}|}_{j=i+1}\exp\left(\frac{\|\mathbf{X}_{i}^{\mathrm{H}}\mathbf{X}_{j}\|^2_\mathrm{F}
    }{\epsilon}\right),\label{objMCD}
\end{align}
where $\epsilon$ is a smoothing constant. 

\subsection{Noncoherent Detection}
Noncoherent detection can be performed without using channel state information (CSI). 
The Grassmann constellation satisfies orthonormality, and each subspace is distinct. 
Utilizing this property in communication allows for codewords that are invariant under multiplication by the channel matrix.

A Grassmann codeword $\mathbf{X}$ with time slot length $T=T_\mathrm{g}$ is transmitted according to the system model \eqref{systemmodel}, where $\mathbf{X}$ satisfies $\mathbf{X}^{\mathrm{H}}\mathbf{X} = \mathbf{I}_M$.
Since $\mathrm{E}[\|\mathbf{X}\|^{2}_{\mathrm{F}}]=M$, to satisfy $\mathrm{E}[\|\mathbf{S}\|^{2}_{\mathrm{F}}]=T_\mathrm{g}$, we set $\mathbf{S}=\sqrt{T_\mathrm{g}/M}\mathbf{X}$.
The generalized likelihood ratio test (GLRT) receiver can estimate the codeword by maximizing the likelihood without CSI and is given by \cite{warrier2002spectrally}
\begin{align}
    \hat{\mathbf{X}}=\underset{\mathbf{X}\in\mathcal{X}}{\mathrm{argmax}}\|\mathbf{Y}^{\mathrm{H}}\mathbf{X}\|^2_{\mathrm{F}},
    \label{GLRT}
\end{align}
where $\hat{\mathbf{X}}$ denotes an estimated symbol. 

\subsection{Noncoherent AMI of Grassmann Communication}
Let $B_{\mathrm{g}}$ be the number of transmission bits of a Grassmann constellation, and noncoherent AMI is calculated by the Monte Carlo simulation of \cite{ngo2020cubeSplit}:
\begin{align}
R_{\mathrm{g}}=\frac{B_{\mathrm{g}}}{T_{\mathrm{g}}}-\frac{1}{T_{\mathrm{g}}|\mathcal{X}|}\mathrm{E}_{\mathbf{H}, \mathbf{V}}\left[\sum_{i=1}^{|\mathcal{X}|}\log_2\left(\frac{\sum_{j=1}^{|\mathcal{X}|}p(\mathbf{Y}_i|\mathbf{X}_j)}{p(\mathbf{Y}_i|\mathbf{X}_i)}\right)\right].\label{AMI_Rg}
\end{align}
In multi-antenna scenarios, the conditional probability $p(\mathbf{Y}_i|\mathbf{X}_j)$ in \eqref{AMI_Rg} can be calculated as
\cite{gohary2009noncoherent,endo2024boosting}
\begin{align}
p(\mathbf{Y}|\mathbf{X})=\frac{\exp(-\|\mathbf{Y}\|_\mathrm{F}^2/\sigma_{\mathrm{v}}^2+\|\mathbf{Y}^\mathrm{H}\mathbf{X}\|_\mathrm{F}^2/\sigma_{\mathrm{v}}^2/\left(1+\sigma_{\mathrm{v}}^2M/T_{\mathrm{g}}\right))}{\pi^{T_{\mathrm{g}}N}\sigma_{\mathrm{v}}^{2T_{\mathrm{g}}N}\left(1+1/\sigma_{\mathrm{v}}^2/M\cdot T_{\mathrm{g}}\right)^{MN}.
}
\end{align}
Then, we express $p(\mathbf{Y}_i|\mathbf{X}_j)/p(\mathbf{Y}_i|\mathbf{X}_i)=\exp(\eta_{i,j})$ in \eqref{AMI_Rg}.
Here, $\eta_{i,j}$ can be expressed as \cite{endo2024boosting}
\begin{align}
\eta_{i,j}=\frac{\|\mathbf{Y}_i^\mathrm{H}\mathbf{X}_j\|_\mathrm{F}^2-\|\mathbf{Y}_i^\mathrm{H}\mathbf{X}_i\|_\mathrm{F}^2}{\sigma_{\mathrm{v}}^2\left(1+\sigma_{\mathrm{v}}^2M/T_{\mathrm{g}}\right)}.\label{eta}
\end{align}
Using $\eta_{i,j}$, we can write \eqref{AMI_Rg} as
\begin{align}
R_{\mathrm{g}}&=\frac{B_{\mathrm{g}}}{T_{\mathrm{g}}}-\frac{1}{T_{\mathrm{g}}|\mathcal{X}|}\mathrm{E}_{\mathbf{H},\mathbf{V}}\left[\sum_{i=1}^{|\mathcal{X}|} \log_2\sum_{j=1}^{|\mathcal{X}|}\exp(\eta_{i,j})\right].\label{AMI_G}
\end{align}

\subsection{Data-Carrying Reference Signal}
The channel can be estimated from the estimated symbol $\hat{\mathbf{X}}$ obtained through noncoherent communication using DC-RS.
Assuming $\mathbf{X}=\hat{\mathbf{X}}$, we can obtain the estimated channel $\hat{\mathbf{H}}$ by multiplying $\sqrt{M/T_{g}}\hat{\mathbf{X}}^{\mathrm{H}}$ from the left side of the received symbol $\mathbf{Y}$, written as \cite{endo2024boosting}
\begin{align}
    \hat{\mathbf{H}}=\sqrt{\frac{M}{T_{g}}}\hat{\mathbf{X}}^{\mathrm{H}}\mathbf{Y}=\hat{\mathbf{X}}^{\mathrm{H}}\mathbf{X}\mathbf{H}+\sigma_{\mathrm{v}}\mathbf{V'}=\mathbf{H}+\sigma_{\mathrm{v}}\mathbf{V'},
    \label{channel_est}
\end{align}
where $\mathbf{V'}=\sqrt{M/T_{g}}\hat{\mathbf{X}}^{\mathrm{H}}\mathbf{V}$.
This enables DC-RS to estimate the channel while simultaneously transmitting data.
When $\hat{\mathbf{X}}\neq\mathbf{X}$, DC-RS cannot estimate the channel accurately, leading to a degradation in channel estimation accuracy compared with conventional RS.

\begin{figure}[tb]
	\centering
	\includegraphics[clip,scale=0.8]{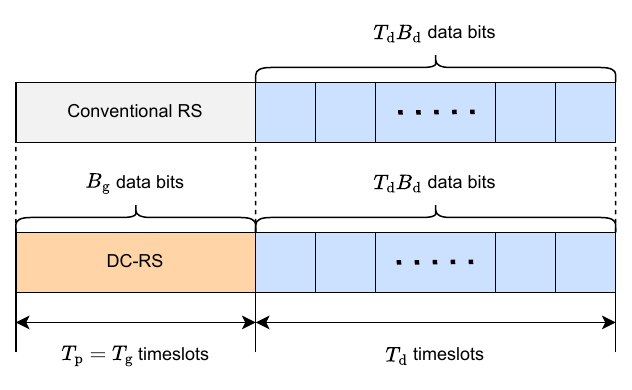}
	\caption{Example of conventional RS and DC-RS structures \cite{endo2024boosting}. \label{fig:RS}}
\end{figure}

An example of communication using conventional RS and DC-RS is shown in Fig.~\ref{fig:RS} \cite{endo2024boosting}.
In this example, an RS is transmitted for $T_\mathrm{p}$ or $T_\mathrm{g}$ time slots, followed by coherent symbols transmitted for $T_\mathrm{d}$ time slots. 
The transmitted $B_{\mathrm{d}}$ bits of coherent symbols are detected on the basis of the channel estimated by the RS.
Although conventional RS carries no information, DC-RS conveys an additional $B_{\mathrm{g}}$ of information.
Therefore, it is possible to increase the amount of information transmitted within a frame; however, since the channel estimation accuracy of conventional RS is superior, an evaluation of the actually conveyed information is necessary.

\subsection{NMSE Minimization by Unitary Matrices}
In \cite{endo2024boosting}, a method was proposed to minimize the NMSE of DC-RSs. 
By considering symbol estimation errors, we can expand $\mathrm{E}[\|\hat{\mathbf{H}}-\mathbf{H}\|_{\mathrm{F}}^2]$ as \cite{endo2024boosting}
\begin{align}
    \mathrm{E}[\| \Hh - \H \|_{\mathrm{F}}^2]
    = \sigma_{\mathrm{v}}^2 \frac{M^2N}{T_{\mathrm{g}}} +
    \frac{2}{|\mathcal{X}|}      \sum^{|\mathcal{X}|-1}_{i=1}\sum^{|\mathcal{X}|}_{j=i+1}
    p_{i, j} \mathrm{E}[\| \D_{i,j} \|_{\mathrm{F}}^2],\label{eq:HohatmH}
\end{align}
where $\D_{i, j} = (\X_j^{\mathrm{H}} \X_i - \I_M) \H$ and $p_{i, j}$ denotes the PEP.
The multiplication of the Grassmann constellation by a unitary matrix does not change their subspace, thus having no effect on the average power or the chordal distance between codewords.
Using this property, we can formulate the NMSE minimization problem as \cite{endo2024boosting}
\begin{equation}
\begin{aligned}
    \minimize_{\{\mathbf{U}_1, \cdots, \mathbf{U}_{|\mathcal{X}|}\}} \quad & \sum^{|\mathcal{X}|-1}_{i=1}\sum^{|\mathcal{X}|}_{j = i + 1}
    \frac{
    \left\|
    \mathbf{I}_M -
    \mathbf{U}_i^{\mathrm{H}}
    \X_{i}^{\mathrm{H}}
    \X_{j}\mathbf{U}_j
    \right\|_{\mathrm{F}}^2}
    {\mathrm{Re}[\det(\I_M - \X_i^{\mathrm{H}}\X_j\X_j^{\mathrm{H}}\X_i)]}
    \\
    \textrm{s.t.} \quad & \mathbf{U}_i \in \mathcal{U}(M),\;\forall\; i=\{1,\cdots,|\mathcal{X}|\}.
\end{aligned}
\label{eq:opti_phase}
\end{equation}

\section{Proposed Total AMI optimization Method}
In this section, we propose an optimization method for the total AMI of communication with DC-RSs. 
We derive objective functions for optimizing the NMSE of estimated channel matrices and noncoherent AMI, and we propose a method that uses a convex combination to optimize these functions simultaneously.

\subsection{Problem Formulation}
We assess the achievable spectral efficiency of communication with DC-RSs, referred to as the total AMI, which is composed of coherent AMI and noncoherent AMI.
According to \eqref{AMI_d} - \eqref{z_ij}, coherent AMI depends solely on SNR and the NMSE of estimated channel matrices.
Hence, the total AMI depends only on a constructed Grassmann codebook $\mathcal{X}$ and SNR. 
Here, we define the total AMI as
\begin{align}
R_{\mathrm{T}}(\mathcal{X},\sigma_{\mathrm{v}}^2)=T_{\mathrm{g}}R_{\mathrm{g}}+T_{\mathrm{d}}R_{\mathrm{d}},
\end{align}
and its maximization problem can be formulated as
\begin{equation}
\begin{aligned}
    \underset{\mathcal{X}=\{\mathbf{X}_1, \cdots, \mathbf{X}_{|\mathcal{X}|}\}}{\mathrm{maximize}}\quad & R_{\mathrm{T}}(\mathcal{X},\sigma_{\mathrm{v}}^2)\
    \\
    \textrm{s.t.} \quad & \mathbf{X}_i\in\mathcal{G}(T_{\mathrm{g}},M),\;\forall\; i=\{1,\cdots,|\mathcal{X}|\}.
\end{aligned}
\end{equation}
To maximize the total AMI, DC-RSs should be optimized for both noncoherent AMI and NMSE. 
However, Monte Carlo simulations are computationally intensive, making it challenging to directly optimize performance metrics such as AMI and NMSE themselves.
Additionally, there is a trade-off between noncoherent AMI and NMSE, and optimizing the total AMI requires a dedicated simultaneous optimization of both metrics.

\subsection{New Upper Bound for NMSE}
As given in \eqref{eq:opti_phase}, the NMSE of estimated channel matrices can be minimized via a unitary transformation without changing the subspace.
That is, the optimization of \eqref{eq:opti_phase} does not change the chordal distance, implying that AMI and SER are maintained.
In contrast, this article optimizes the constellation itself to maximize the total AMI.
By optimizing the relationship between subspaces, it is possible to achieve a lower NMSE.
If we use \eqref{eq:opti_phase} as an objective function to minimize the NMSE of estimated channel matrices by optimizing the constellation itself, the minimization problem can be reformulated into
\begin{equation}
\begin{aligned}
    \underset{\mathcal{X}=\{\mathbf{X}_1, \cdots, \mathbf{X}_{|\mathcal{X}|}\}}{\mathrm{minimize}} \quad & \sum^{|\mathcal{X}|-1}_{i=1}\sum^{|\mathcal{X}|}_{j = i + 1}
    \frac{
    \left\|
    \mathbf{I}_M -
    \X_{i}^{\mathrm{H}}
    \X_{j}
    \right\|_{\mathrm{F}}^2}
    {\mathrm{Re}[\det(\I_M - \X_i^{\mathrm{H}}\X_j\X_j^{\mathrm{H}}\X_i)]}
    \\
    \textrm{s.t.} \quad & \mathbf{X}_i\in\mathcal{G}(T_{\mathrm{g}},M),\;\forall\; i=\{1,\cdots,|\mathcal{X}|\}.
\end{aligned}
\label{NMSEminimization_nonunitary}
\end{equation}

If we were to use \eqref{NMSEminimization_nonunitary} as an objective function, the PEP would be minimized at the same time because \eqref{NMSEminimization_nonunitary} includes weighted coefficients of PEPs.
This is not an issue if only NMSE minimization is considered, but our method simultaneously maximize noncoherent AMI.
The problem here is that there is a positive correlation between PEP and noncoherent AMI, whereas NMSE and noncoherent AMI have a negative correlation. 
Hence, the PEP minimization leads to improvements in both NMSE and noncoherent AMI within the convex combination, thereby reducing the degrees of freedom for optimization.
By excluding the weighted coefficients of PEPs from the NMSE objective function, we can optimize NMSE with greater freedom, yielding a constellation that has better total AMI.

To avoid the above issue, we instead use an upper bound of PEP. 
The upper bound of \eqref{eq:HohatmH} is transformed into
\begin{align}
    &\mathrm{E}[\| \Hh - \H \|_{\mathrm{F}}^2]\nonumber
    \\
    &\leq\sigma_{\mathrm{v}}^2 \frac{M^2N}{T_{\mathrm{g}}} +
    \frac{2}{|\mathcal{X}|}\left(\underset{i\neq j}{\mathrm{max}}\ p_{i,j}\right)      \sum^{|\mathcal{X}|-1}_{i=1}\sum^{|\mathcal{X}|}_{j=i+1}
    \mathrm{E}[\| \D_{i,j} \|_{\mathrm{F}}^2] .\label{HhH_upper}
\end{align}
Thus, we can define a new objective function $f_{\mathrm{NMSE}}$ as
\begin{align}
    f_{\mathrm{NMSE}}:=\sum^{|\mathcal{X}|-1}_{i=1}\sum^{|\mathcal{X}|}_{j = i + 1}\left\|\mathbf{I}_M -\X_{i}^{\mathrm{H}}\X_{j}\right\|_{\mathrm{F}}^2,
\end{align}
where the minimization of $f_{\mathrm{NMSE}}$ is equivalent to that of \eqref{HhH_upper}.
In this paper, we propose $f_{\mathrm{NMSE}}$ as the objective function for minimizing NMSE in the problem of maximizing the total AMI. 

When only $f_{\mathrm{NMSE}}$ is minimized, the NMSE performance of the optimized constellation is equivalent to that of the training method.
However, this occurs only when all codewords satisfy $\X_{i}^{\mathrm{H}}\X_{j}=\mathbf{I}_M$, meaning that all codewords are identical.
This is equivalent to the receiver having prior knowledge of the transmitted codeword, resulting in no information.
Therefore, to maximize the total AMI of DC-RSs, it is necessary to consider not only the NMSE but also the noncoherent AMI.

\subsection{New Lower Bound for Noncoherent AMI}
Maximizing the MCD leads to the maximization of noncoherent AMI.
However, in this paper, we consider the simultaneous optimization of two performance metrics in a trade-off relationship, namely, NMSE and noncoherent AMI; in other words, we attempt to optimize both objectives. 
Accordingly, we derive a new objective function that is related to the noncoherent AMI, leading to a flexible maximization of the total AMI.

The noncoherent AMI can be obtained from \eqref{AMI_G}, but transforming the Monte Carlo integration in closed form is a challenging task.
Here, we instead derive a lower bound for the noncoherent AMI and maximize it. 
First, by applying Jensen's inequality to \eqref{AMI_G}, we obtain the lower bound of the noncoherent AMI that requires Monte Carlo simulations, which is given by
\begin{align}
R_{\mathrm{g}}&\geq\frac{B_{\mathrm{g}}}{T_{\mathrm{g}}}-\frac{1}{T_{\mathrm{g}}|\mathcal{X}|}\log_2\sum_{i=1}^{|\mathcal{X}|} \sum_{j=1}^{|\mathcal{X}|}\mathrm{E}_{\mathbf{H},\mathbf{V}}\left[\exp(\eta_{i,j})\right].
\end{align}
In particular, by using the Maclaurin expansion at low SNRs, we can expand $\mathrm{E}_{\mathbf{H},\mathbf{V}}\left[\exp(\eta_{i,j})\right]$ as
\begin{align}
    \mathrm{E}_{\mathbf{H},\mathbf{V}}\left[\exp(\eta_{i,j})\right]\simeq\mathrm{E}_{\mathbf{H},\mathbf{V}}\left[1+\eta_{i,j}\right],
\end{align}
where
\begin{align}
    \mathrm{E}_{\mathbf{H},\mathbf{V}}\left[\eta_{i,j}\right]
    =\frac{\mathrm{E}_{\mathbf{H},\mathbf{V}}\left[\|\mathbf{Y}_i^\mathrm{H}\mathbf{X}_j\|_\mathrm{F}^2\right]-\mathrm{E}_{\mathbf{H},\mathbf{V}}\left[\|\mathbf{Y}_i^\mathrm{H}\mathbf{X}_i\|_\mathrm{F}^2\right]}{\sigma_{\mathrm{v}}^2\left(1+\sigma_{\mathrm{v}}^2M/T_{\mathrm{g}}\right)},\label{average_norm}
\end{align}
and $\mathrm{E}_{\mathbf{H},\mathbf{V}}\left[\|\mathbf{Y}_i^\mathrm{H}\mathbf{X}_j\|_\mathrm{F}^2\right]$ can be written as
\begin{align}
    &\mathrm{E}_{\mathbf{H},\mathbf{V}}\left[\|\mathbf{Y}_i^\mathrm{H}\mathbf{X}_j\|_\mathrm{F}^2\right]\nonumber\\
    &=\mathrm{E}_{\mathbf{H},\mathbf{V}}\left[\|(\mathbf{X}_i\mathbf{H}+\mathbf{V})^\mathrm{H}\mathbf{X}_j\|_\mathrm{F}^2\right]\nonumber\nonumber\\
    &=\mathrm{E}_{\mathbf{H},\mathbf{V}}\left[\mathrm{Tr}\left\{\left((\mathbf{X}_i\mathbf{H}+\mathbf{V})^\mathrm{H}\mathbf{X}_j\right)\left((\mathbf{X}_i\mathbf{H}+\mathbf{V})^\mathrm{H}\mathbf{X}_j\right)^\mathrm{H}\right\}\right]\nonumber\\
    &=\mathrm{E}_{\mathbf{H},\mathbf{V}}\left[\mathrm{Tr}\left\{(\mathbf{X}_i\mathbf{H}+\mathbf{V})^\mathrm{H}\mathbf{X}_j\mathbf{X}_j^\mathrm{H}(\mathbf{X}_i\mathbf{H}+\mathbf{V})\right\}\right]\nonumber\\
    &\ \ \ \ \ \ \ \ \ \ \ \ \ \ \ \ \ \ \ \ \ \ \ \left.\left.+\mathbf{V}^\mathrm{H}\mathbf{X}_j\mathbf{X}_j^\mathrm{H}\mathbf{X}_i\mathbf{H}+\mathbf{V}^\mathrm{H}\mathbf{X}_j\mathbf{X}_j^\mathrm{H}\mathbf{V}
    \right\}\right]\nonumber\\
    &=E_1+E_2+E_3+E_4.
\end{align}
We calculate the expectation of each term.
For the first term, since the expectation of each element of $\mathbf{H}\mathbf{H}^\mathrm{H}$ is $1$ under the assumption of Rayleigh fading, $E_1$ is given by
\begin{align}
    E_1&=\mathrm{E}_{\mathbf{H},\mathbf{V}}\left[\mathrm{Tr}\left\{\mathbf{H}^\mathrm{H}\mathbf{X}_i^\mathrm{H}\mathbf{X}_j\mathbf{X}_j^\mathrm{H}\mathbf{X}_i\mathbf{H}\right\}\right]\nonumber\\
    &=\mathrm{Tr}\left\{\mathbf{X}_i^\mathrm{H}\mathbf{X}_j\mathbf{X}_j^\mathrm{H}\mathbf{X}_i\right\}\nonumber\\
    &=\|\mathbf{X}_i^\mathrm{H}\mathbf{X}_j\|^2_\mathrm{F}.
\end{align}
Next, for the second and third terms, since $\mathbf{V}$ and $\mathbf{H}$ are independent, $E_2$ and $E_3$ are respectively given by
\begin{align}
    E_2=\mathrm{E}_{\mathbf{H},\mathbf{V}}\left[\mathrm{Tr}\left\{\mathbf{H}^\mathrm{H}\mathbf{X}_i^\mathrm{H}\mathbf{X}_j\mathbf{X}_j^\mathrm{H}\mathbf{V}\right\}\right]=0,
\end{align}
and
\begin{align}
    E_3=\mathrm{E}_{\mathbf{H},\mathbf{V}}\left[\mathrm{Tr}\left\{\mathbf{V}^\mathrm{H}\mathbf{X}_j\mathbf{X}_j^\mathrm{H}\mathbf{X}_i\mathbf{H}\right\}\right]=0.
\end{align}
Finally, for the fourth term, since the expectation of each element of $\mathbf{V}\mathbf{V}^\mathrm{H}$ is $\sigma_\mathrm{v}^2$, $E_4$ is given by
\begin{align}
    E_4&=\mathrm{E}_{\mathbf{H},\mathbf{V}}\left[\mathrm{Tr}\left\{\mathbf{V}^\mathrm{H}\mathbf{X}_j\mathbf{X}_j^\mathrm{H}\mathbf{V}\right\}\right]\nonumber\\
    &=\sigma_\mathrm{v}^2\|\mathbf{X}_j\|^2_\mathrm{F}\nonumber\\
    &=M\sigma_\mathrm{v}^2.
\end{align}
Consequently, $\mathrm{E}_{\mathbf{H},\mathbf{V}}\left[\eta_{i,j}\right]$ of \eqref{average_norm} can be simplified into 
\begin{align}
    \mathrm{E}_{\mathbf{H},\mathbf{V}}\left[\eta_{i,j}\right]
    &=\frac{\|\mathbf{X}_i^\mathrm{H}\mathbf{X}_j\|^2_\mathrm{F}+M\sigma_{\mathrm{v}}^2-(\|\mathbf{X}_i^\mathrm{H}\mathbf{X}_i\|^2_\mathrm{F}+M\sigma_{\mathrm{v}}^2)}{\sigma_{\mathrm{v}}^2\left(1+\sigma_{\mathrm{v}}^2M/T_{\mathrm{g}}\right)}\nonumber\\
    &=\frac{\|\mathbf{X}_i^\mathrm{H}\mathbf{X}_j\|^2_\mathrm{F}-\|\mathbf{X}_i^\mathrm{H}\mathbf{X}_i\|^2_\mathrm{F}}{\sigma_{\mathrm{v}}^2\left(1+\sigma_{\mathrm{v}}^2M/T_{\mathrm{g}}\right)}\nonumber\\
    &=\frac{\|\mathbf{X}_i^\mathrm{H}\mathbf{X}_j\|^2_\mathrm{F}-M}{\sigma_{\mathrm{v}}^2\left(1+\sigma_{\mathrm{v}}^2M/T_{\mathrm{g}}\right)},
\end{align}
and the lower bound of noncoherent AMI is written as
\begin{align}
    R_{\mathrm{g}}&\geq\frac{B_{\mathrm{g}}}{T_{\mathrm{g}}}-\frac{1}{T_{\mathrm{g}}|\mathcal{X}|}\log_2\left(|\mathcal{X}|^2+\sum_{i=1}^{|\mathcal{X}|} \sum_{j=1}^{|\mathcal{X}|}\frac{\|\mathbf{X}_i^\mathrm{H}\mathbf{X}_j\|^2_\mathrm{F}-M}{\sigma_{\mathrm{v}}^2\left(1+\sigma_{\mathrm{v}}^2M/T_{\mathrm{g}}\right)}\right).\label{AMI_under}
\end{align}
Therefore, by using the symmetric relationship of $\mathbf{X}_i$ and $\mathbf{X}_j$, maximizing this lower bound is equivalent to minimizing a new objective function:
\begin{align}
    f_{\mathrm{SCD}}:=\sum^{|\mathcal{X}|-1}_{i=1}\sum^{|\mathcal{X}|}_{j = i + 1}\|\mathbf{X}_i^\mathrm{H}\mathbf{X}_j\|^2_\mathrm{F},\label{upp}
\end{align}
which is referred to as the sum of chordal distances (SCD).

\begin{figure}[tb]
	\centering
    \subfigure[Conventional $f_{\mathrm{MCD}}$.]{
		\includegraphics[clip, scale=0.72]{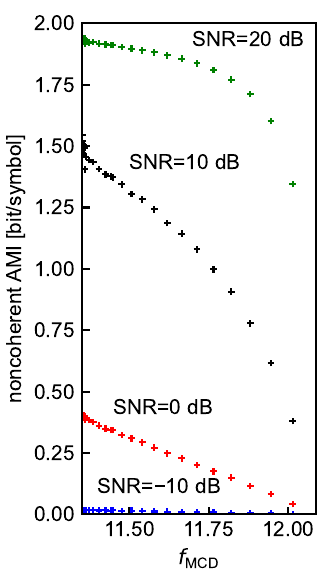}
	}
    \subfigure[Proposed $f_{\mathrm{SCD}}$.]{
		\includegraphics[clip, scale=0.72]{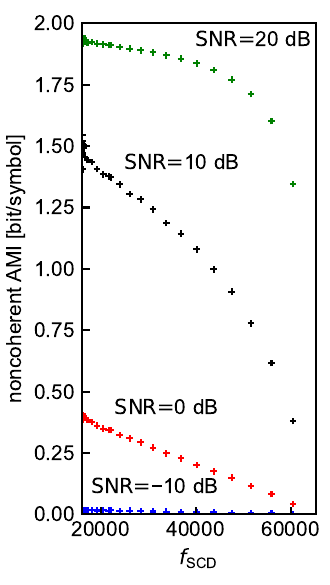}
	}
	\caption{Relationships between noncoherent AMI and objective functions.\label{fig:fvsAMI}}
\end{figure}
The relationship between the noncoherent AMI and the objective functions $f_{\mathrm{MCD}}$ and $f_{\mathrm{SCD}}$ is shown in Fig. \ref{fig:fvsAMI}.
Here, we set parameters as $(M, T_{\mathrm{g}}, B_{\mathrm{g}}) = (1, 4, 8)$, and SNRs as $-10, 0,10$ and $20\ \mathrm{dB}$.
As shown in Fig. \ref{fig:fvsAMI}, both objective functions are sharply correlated with the noncoherent AMI in a wide range of SNRs.
Although $f_{\mathrm{MCD}}$ has conventionally been used as a standard objective function to maximize the noncoherent AMI, we combine $f_{\mathrm{NMSE}}$ to maximize not only the noncoherent AMI but the total AMI. Thus, we demonstrate the advantages of using $f_{\mathrm{SCD}}$ over $f_{\mathrm{MCD}}$ via the performance comparisons in Section $\mathrm{V}$.

\subsection{Total AMI Optimization Method}
In the previous sections, the new objective functions $f_{\mathrm{NMSE}}$ and $f_{\mathrm{SCD}}$ were obtained.
As mentioned earlier, there is a trade-off between the NMSE of estimated channel matrices and noncoherent AMI.
Thus, to maximize the total AMI, these objective functions must be optimized in a dedicated manner.
For this multi-objective optimization problem, we propose optimizing a convex combination of objective functions to obtain constellations on the Pareto front for the NMSE and noncoherent AMI.
Letting $f_{\mathrm{C}}$ and $f_{\mathrm{R}}$ be the objective functions for the NMSE of estimated channel matrices and noncoherent AMI, which is an upper bound for effective data rate, respectively, the total AMI maximization problem for a given ratio $\alpha \in [0,1]$ can be represented by the minimization problem as follows: 
\begin{align}
    \mathcal{X}^\ast=\underset{\mathcal{X}}{\mathrm{argmin}}
    \quad
    \alpha f_{\mathrm{C}}+(1-\alpha)f_{\mathrm{R}}.\label{object_function}
\end{align}
In our comparisons, we use $f_{\mathrm{C}}$ as $f_{\mathrm{NMSE}}$ and evaluate the difference between $f_{\mathrm{R}}=f_{\mathrm{MCD}}$ and $f_{\mathrm{R}}=f_{\mathrm{SCD}}$.
In this minimization problem, the optimal value of $\alpha$ depends on parameters such as the number of time slots $T_\mathrm{g}$ and transmission bits $B_\mathrm{g}$.
For example, in the slot configuration given in Fig.~\ref{fig:RS}, when the proportion of bits in the coherent code is larger than that in the noncoherent code, the contribution of NMSE to the total AMI becomes greater, making a relatively large $\alpha$ optimal.
Therefore, we also have to optimize $\alpha$ for a parameter set.
For each $\alpha$, we perform manifold optimization on \eqref{object_function} using Mopt.
The total AMI is then calculated for the constellation obtained through optimization, and this serves as the score for each $\alpha$. 
Hence, the problem of maximizing the total AMI with respect to $\alpha$ at a given SNR is expressed by
\begin{equation}
\begin{aligned}
    \underset{\alpha}{\mathrm{maximize}}\quad&
    R_{\mathrm{T}}(\mathcal{X}^\ast,\sigma_{\mathrm{v}}^2)\\
    \textrm{s.t.}\quad & \alpha \in [0,1].
\end{aligned}
\label{object_function_alpha}
\end{equation}

To efficiently search for the optimal $\alpha$, we employ an algorithm known as the tree-structured Parzen estimator (TPE) \cite{bergstra2011algorithms}, a type of Bayesian optimization. 
In TPE, the results of the objective function are divided into a set $\mathcal{L}$ of $\alpha$ that yield good scores and a set $\mathcal{G}$ of those that do not, then the probability density functions $l(\alpha)$ for $\mathcal{L}$ and $g(\alpha)$ for $\mathcal{G}$ are constructed by kernel density estimation.
From these probability density functions, the $\alpha$ of the next trial is selected as follows:

\begin{align}
    \alpha=\underset{\alpha}{\mathrm{argmax}}\frac{l(\alpha)}{g(\alpha)}.\label{TPEalg}
\end{align}

The optimization algorithm is summarized in Algorithm \ref{algorithm_total}.
In line 2, the probability density functions are constructed by kernel density estimation.
In line 3, for the first few trials, random values of $\alpha \in [0,1]$ are selected to construct the probability density functions.
For each trial, \eqref{object_function} is optimized on selected $\alpha$, and the total AMI of the obtained constellation is the score.
By repeating this trial, we can obtain the optimal $\alpha$ using the TPE algorithm.

\begin{algorithm}[tb]
    \caption{Proposed total AMI optimization algorithm.}
    \begin{algorithmic}[1]
        \renewcommand{\algorithmicrequire}{\textbf{Input:}}
        \renewcommand{\algorithmicensure}{\textbf{Output:}}
        \REQUIRE Objective functions $f_{\mathrm{C}}$, $f_{\mathrm{R}}$, and $R_{\mathrm{T}}(\mathcal{X}^\ast,\sigma_{\mathrm{v}}^2)$, SNR $\sigma_{\mathrm{v}}^2$, the number of trials $N_o$
        \FOR{each trial $t = 1$ to $N_o$}
            \STATE Construct $l(\alpha)$ and $g(\alpha)$ from prior trial results.
            \STATE Select $\alpha$ by calculating \eqref{TPEalg}.
            \STATE Optimization by calculating \eqref{object_function} using Mopt yields $\mathcal{X}^\ast$.
            \STATE Calculate total AMI $R_{\mathrm{T}}(\mathcal{X}^\ast,\sigma_{\mathrm{v}}^2)$.
            \STATE Record the score for $\alpha$ as $R_{\mathrm{T}} (\mathcal{X}^\ast,\sigma_{\mathrm{v}}^2)$ .
        \ENDFOR 
        \STATE \textbf{return} $\alpha$ and corresponding $\mathcal{X}^\ast$ that maximizes $R_{\mathrm{T}}(\mathcal{X}^\ast,\sigma_{\mathrm{v}}^2)$.
        \ENSURE Optimal $\alpha$ and $\mathcal{X}^\ast$
    \end{algorithmic}
    \label{algorithm_total}
\end{algorithm}

\section{Performance Comparisons}
\label{sec:comp}
In this section, we demonstrate the performance advantage of the constellation obtained through the proposed optimization method. 
Here, the numbers of time slots and bits for the Grassmann constellation were set to $T_{\mathrm{g}}=4$ and $B_{\mathrm{g}}=8$, respectively.
In the calculation of the total AMI for the optimization, SNR was set to $10\ \mathrm{dB}$ and the number of time slots for the data symbols was set to $T_{\mathrm{d}}=10$ where each data symbol is 16QAM.
For the construction by using our proposed optimization method, we used Optuna \cite{akiba2019optuna}, which supports TPE, to search for the optimal value of $\alpha$.

\subsection{Advantages of $f_{\mathrm{SCD}}$ over $f_{\mathrm{MCD}}$}
First, we compare the optimization objective functions $f_{\mathrm{R}}=f_{\mathrm{MCD}}$ and $f_{\mathrm{SCD}}$ in \eqref{object_function}. 
For the optimization of the constellation in this comparison, the numbers of transmit and receive antennas were set to $(M,N)=(1,1)$. 
Accordingly, $B_{\mathrm{d}}=40$, resulting in a total of $B_{\mathrm{g}}+B_{\mathrm{d}}=48$ bits transmitted. 
Here, in the case of $(f_{\mathrm{C}},f_{\mathrm{R}})=(f_{\mathrm{NMSE}},f_{\mathrm{MCD}})$, we set $\epsilon$ of \eqref{objMCD} as $10^{-4}$ to ensure that the contributions of each objective function are balanced, as there is a large difference between the maximum values of $f_{\mathrm{NMSE}}$ and $f_{\mathrm{MCD}}$.

\begin{figure}[tb]
	\centering
    \includegraphics[clip, scale=0.70]{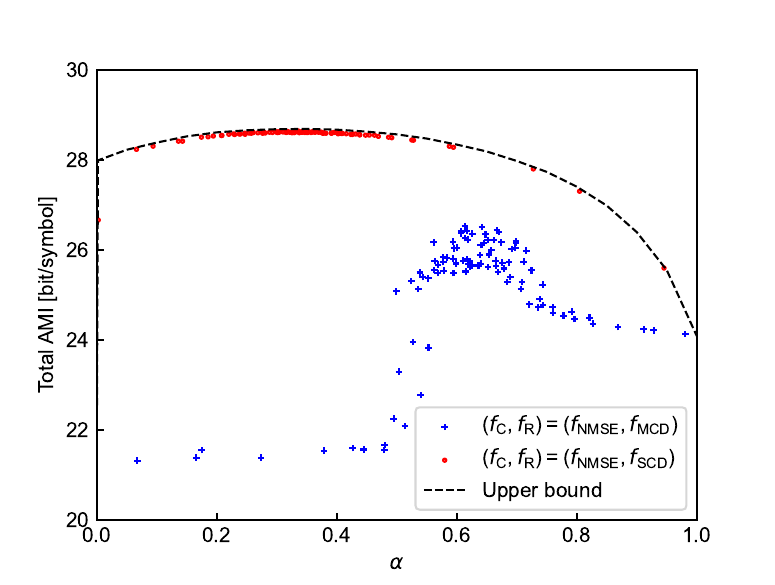}
	\caption{Results of total AMI optimization with respect to $\alpha$.\label{fig:ALLAMIalp}}
\end{figure}

First, Fig.~\ref{fig:ALLAMIalp} shows the results of the total AMI comparison of the constellations, which were obtained through each trial of the proposed algorithm.
An upper bound, which was obtained using the constellations optimized over a sufficient time by selecting $\alpha$ at equal intervals, is shown in the figure.
As shown in Fig.~\ref{fig:ALLAMIalp}, the results obtained using $f_{\mathrm{SCD}}$ were better than those obtained using $f_{\mathrm{MCD}}$ in terms of the total AMI. 
Furthermore, the search using $f_{\mathrm{SCD}}$ was stable and demonstrated the effective exploration of $\alpha$. 
The reason why optimization using $f_{\mathrm{MCD}}$ was less effective is demonstrated in the next performance comparison.

\begin{figure}[tb]
	\centering
    \includegraphics[clip, scale=0.70]{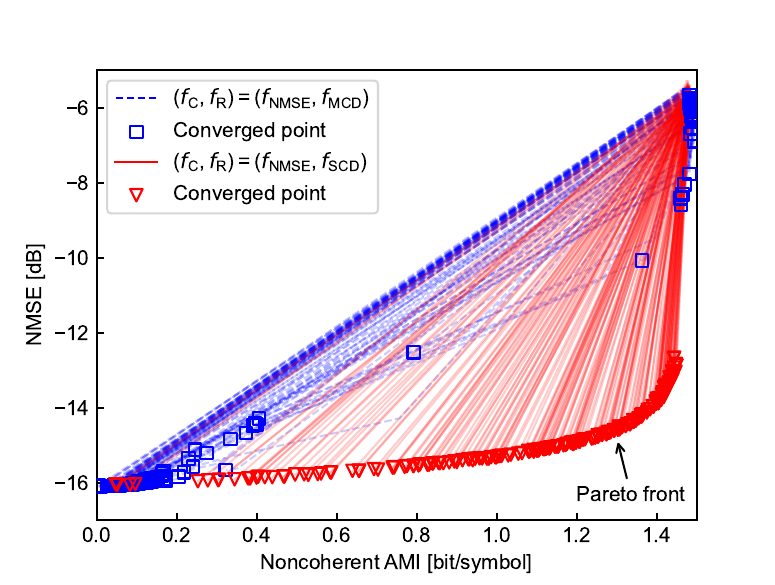}
	\caption{Optimization progress of noncoherent AMI and NMSE.\label{fig:RgNMSEcomp}}
\end{figure}

In Fig.~\ref{fig:RgNMSEcomp}, we compared the progress of optimization in terms of noncoherent AMI and the NMSE of the estimated channel matrices. 
For each constellation, the optimization was performed using random $\alpha$ values, and the optimization trajectories and convergence points were plotted.
For optimization using $f_{\mathrm{SCD}}$, it can be observed that constellations were obtained on the Pareto front with respect to the trade-off between noncoherent AMI and NMSE. 
In contrast, with $f_{\mathrm{MCD}}$, the optimization either barely progressed or became excessively biased towards NMSE, resulting in a polarized outcome. 
Thus, Pareto-optimal constellations were rarely obtained, making $f_{\mathrm{MCD}}$ unsuitable as an optimization objective function.

\subsection{Comparisons of Chordal Distances}
\begin{figure*}[tb]
	\centering
	\subfigure[Exp-Map.]{
		\includegraphics[clip, scale=0.63]{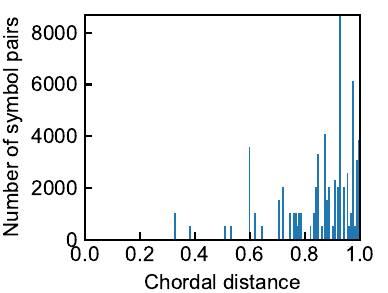}
	}
	\subfigure[Cube-Split.]{
		\includegraphics[clip, scale=0.63]{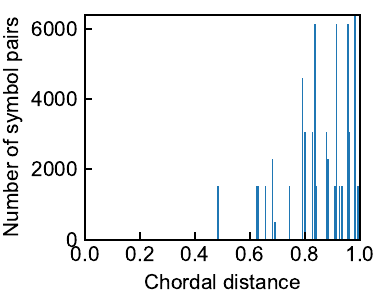}
	}
    \subfigure[Mopt-MCD.]{
		\includegraphics[clip, scale=0.63]{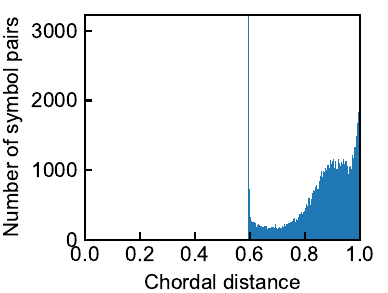}
	}
    \subfigure[Proposed constellation.]{
		\includegraphics[clip, scale=0.63]{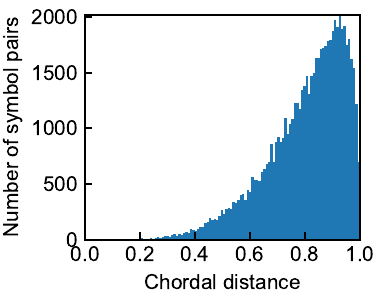}
	}
	\caption{Distribution of chordal distances with $(M, T_{\mathrm{g}}, B_{\mathrm{g}}) = (1, 4, 8)$.\label{fig:DCH}}
\end{figure*}

On the basis of the above results, we will now present the performance of the proposed constellation, which was obtained through optimization using the proposed objective function $f_{\mathrm{SCD}}$.
Here, we compare the chordal distance of the proposed constellation with those of the conventional Grassmann constellations, such as Exp-Map, Cube-Split, and Mopt-MCD, which maximizes the MCD. 
The chordal distance distributions of the proposed constellation and conventional constellations for $(M, T_{\mathrm{g}}, B_{\mathrm{g}}) = (1, 4, 8)$ are shown in Fig.~\ref{fig:DCH}, where Exp-Map was constructed by combining different bits such as $B_{\mathrm{g}}=4+2+2$.
Although algebraically constructed constellations such as Exp-Map and Cube-Split had discrete distributions, the other constellations, which were obtained through numerical optimization, exhibited continuous distributions. 
The proposed constellation, although inferior to Cube-Split and Mopt-MCD in terms of the minimum chordal distance, showed similar peaks to Exp-Map, which excels in channel estimation accuracy, suggesting that optimization including NMSE had been achieved.

\begin{figure}[h]
	\centering
    \includegraphics[clip, scale=0.70]{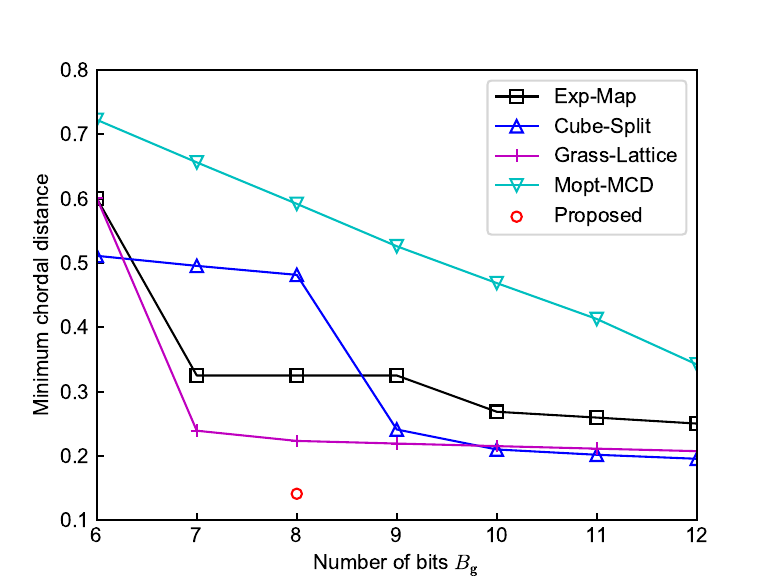}
	\caption{MCD comparison for increasing $B_{\mathbf{g}}$ with $(M,T_{\mathrm{g}})=(1,4)$.\label{fig:compMCD}}
\end{figure}

We also evaluated the MCD for each $B_{\mathrm{g}}$ with the number of time slots fixed at $T_{\mathrm{g}}=4$, as shown in Fig.~\ref{fig:compMCD}.
Here, we compared the constellations obtained through algebraic constructions: Exp-Map, Cube-Split, and Grass-Lattice. 
For Exp-Map and Grass-Lattice, $B_{\mathrm{g}}=6$ and $12$ are the original configurations, whereas for Cube-Split, $B_{\mathrm{g}}=8$ is the original configuration; the other constellations were constructed by combining different bits. 
Thus, for $(M, T_{\mathrm{g}}, B_{\mathrm{g}}) = (1, 4, 8)$, Cube-Split achieved good MCD. 
Grass-Lattice is expected to perform better with larger values of $T_{\mathrm{g}}$. 
As shown in the figure, the constellations derived through algebraic constructions varied in MCD performance based on parameters, with the corresponding performance changed. 
By contrast, Mopt-MCD, which maximizes MCD through numerical optimization, consistently achieved optimal performance, suggesting flexibility across parameters. 
As a reference, we also plotted the MCD of the proposed constellation for $B_{\mathrm{g}}=8$.
It performed bad in terms of MCD than the others because it was optimized to maximize the total AMI rather than MCD. 
To investigate the advantages of the proposed constellation, we evaluated how much channel estimation accuracy and total AMI were improved in the following simulations.

\subsection{Comparisons of NMSE and total AMI}
Here, we compare the NMSE of estimated channel matrices and the total AMI of the constellation obtained by the proposed optimization method with those of the conventional Grassmann constellations and training method. 
The parameters were set to $(T_{\mathrm{g}}, B_{\mathrm{g}}) = (4, 8)$, and comparisons were made for $(M, N) = (1, 1)$ and $(M, N) = (2, 2)$. 
In addition to Mopt-MCD, we also compared a constellation further optimized using unitary matrices, as described in \eqref{eq:opti_phase}.  
We refer to this conventionally optimized constellation \cite{endo2024boosting} as Mopt-NMSE.
For the training method, we set $T_{\mathrm{p}}=4$, used QPSK symbols for the RS, and calculated the total AMI as $T_{\mathrm{d}}R_{\mathrm{d}}$. 
Additionally, in the case of $(M, N) = (2, 2)$, the number of bits for the coherent codes was $B_{\mathrm{d}}=80$, meaning that the proportion of information contained in the coherent codes became larger than that of $(M, N) = (1, 1)$.

\begin{figure}[tb]
	\centering
	\subfigure[$(M,N)=(1,1)$.]{
		\includegraphics[clip, scale=0.70]{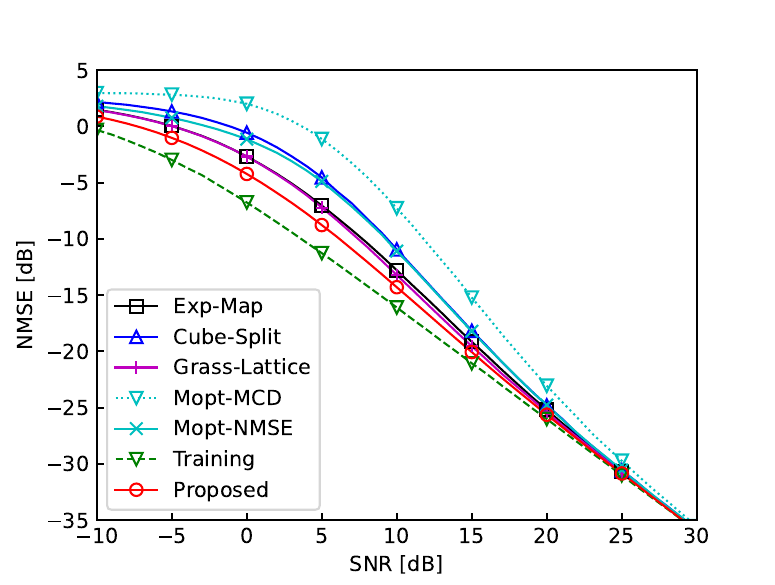}
	}
	\subfigure[$(M,N)=(2,2)$.]{
		\includegraphics[clip, scale=0.70]{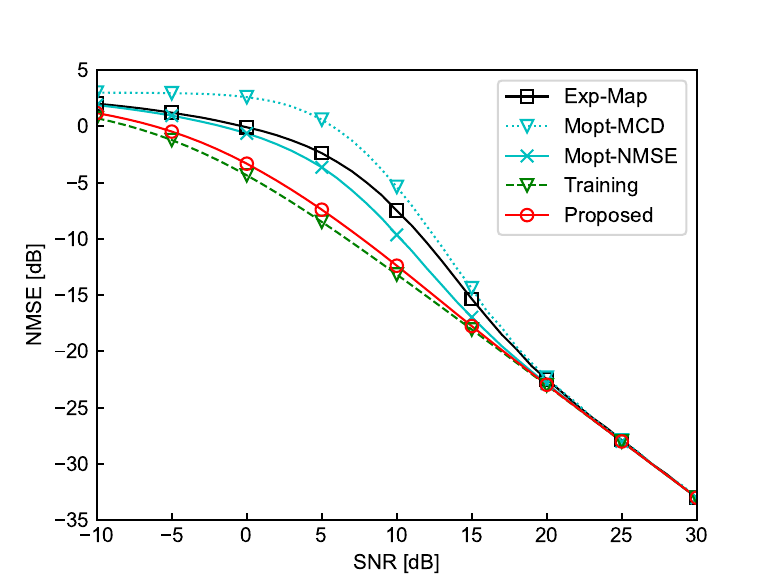}
	}
	\caption{NMSE comparison with $(T_{\mathrm{g}},B_{\mathrm{g}})=(4,8)$.\label{fig:NMSEcomp}}
\end{figure}

First, the NMSE comparison results are shown in Fig.~\ref{fig:NMSEcomp}.
The results indicated that the proposed constellation significantly reduced NMSE compared with the conventional constellations. 
Unlike conventional unitary matrix optimization, the proposed method optimized the subspace itself, achieving a higher channel estimation accuracy than any of the conventional constellations.
Total AMI requires both channel estimation accuracy and noncoherent AMI, and we next show the extent to which this NMSE optimization improves the total AMI.

\begin{figure}[tb]
	\centering
	\subfigure[$(M,N)=(1,1)$.]{
		\includegraphics[clip, scale=0.70]{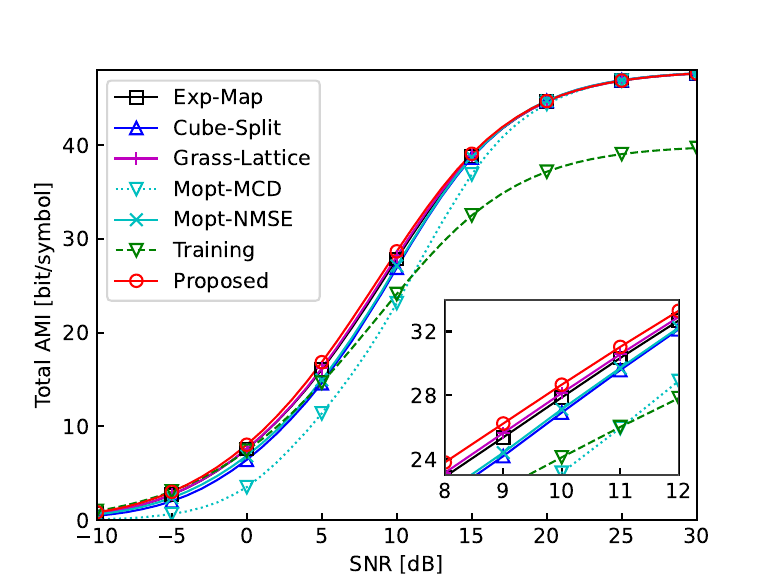}
	}
	\subfigure[$(M,N)=(2,2)$.]{
		\includegraphics[clip, scale=0.70]{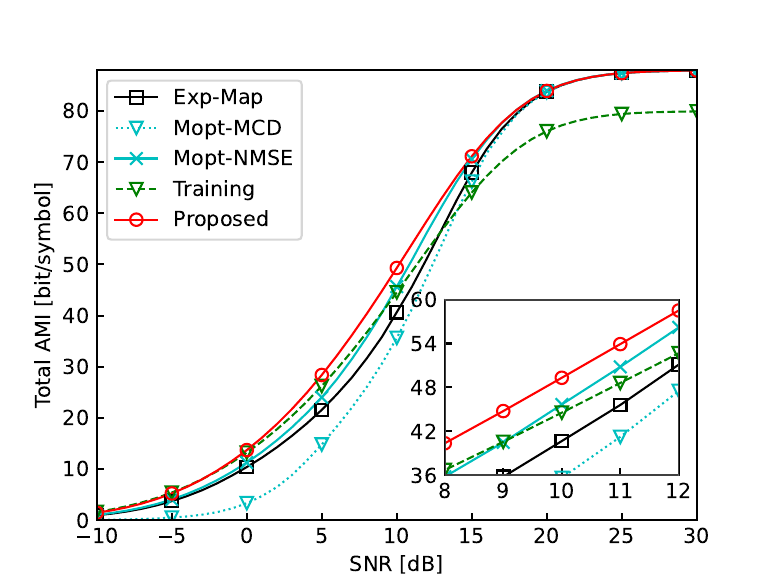}
	}
	\caption{Total AMI comparison with $(T_{\mathrm{g}},B_{\mathrm{g}})=(4,8).\label{fig:TotalAMIcomp}$}
\end{figure}

Next, the total AMI comparison results are shown in Fig.~\ref{fig:TotalAMIcomp}.
These results indicated that the proposed constellation achieved superior performance across all SNRs compared with the conventional constellations. 
Although the conventional constellations failed to achieve improved spectral efficiency over training methods at middle SNRs, the proposed constellation surpassed the performance of the training method. 
Moreover, the proposed constellation achieved performance equal to or better than the conventional constellations at high SNRs. 
In particular, for $(M, N) = (2, 2)$, the improvement at middle SNRs was substantial. 
With an increased number of antennas, the contribution of coherent AMI to the total AMI increased, which heightened the impact of NMSE and resulted in a better total AMI than those obtained using the conventional constellations. 
Therefore, the proposed method can be considered an effective approach to improving spectral efficiency, especially when the proportion of the coherent codes increases.

\textbf{\begin{figure}[h]
	\centering
    \includegraphics[clip, scale=0.70]{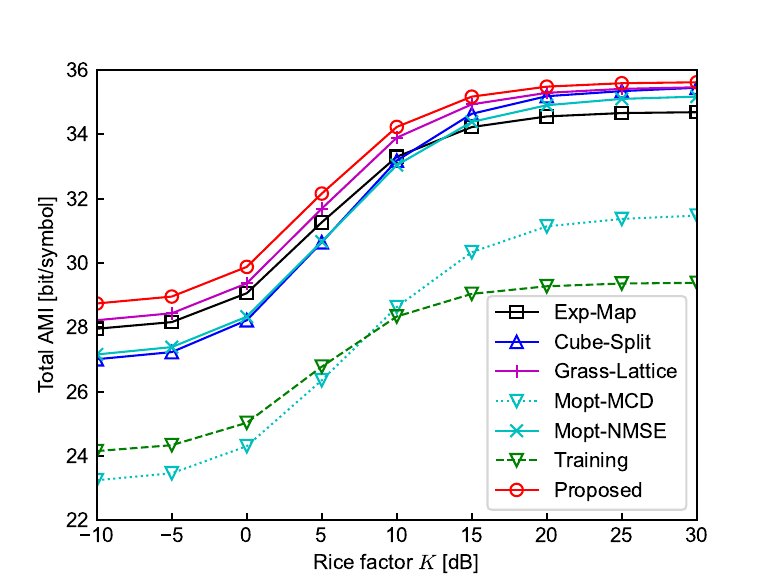}
	\caption{Total AMI comparison upon increasing $K$-factor in Rician channel.\label{fig:Rician}}
\end{figure}}
Finally, Fig. \ref{fig:Rician} shows the comparison results of the total AMI in the Rician channel model $\mathbf{Y}=\mathbf{H^\prime}\mathbf{X}+\sigma_{\mathrm{v}}\mathbf{V}$, where $\mathbf{H}^\prime=\sqrt{K/(K+1)}\cdot\exp{(-2\pi j\cdot d/\lambda)}+\sqrt{1/(K+1)}\mathbf{H}$ in the case of $5\mathrm{GHz}$ carrier frequency.
Here, we set $d=100\mathrm{m}$ and $\lambda=0.06\mathrm{m}$, and the $K$-factor of the Rician channel model is varied from $-10$ to $30\ \mathrm{dB}$.
The parameters were set as $(M, N) = (1, 1)$ for simplicity, and the constellation obtained by the proposed method is the one optimized in the Rayleigh fading channel model.
The comparison was performed with an SNR of $10\ \mathrm{dB}$.
This figure shows that the proposed constellation showed the best performance regardless of the $K$-factor of the Rician channel model, whereas the performance of the other constellations and training method varies with the $K$-factor.
We confirmed that the same trends were observed for the parameters of M, N, and SNR.
Therefore, our proposed constellation can be effective for any channel model in general.

\section{Conclusions}
\label{sec:conc}
In this paper, we proposed a design method for the Grassmann constellation via a numerical optimization, which simultaneously optimizes channel estimation accuracy and data transmission. 
In the numerical optimization, we derived an upper bound for the NMSE of estimated channel matrices and a lower bound for noncoherent AMI, and simultaneously optimized them by using a Bayesian optimization technique. 
The proposed objective functions demonstrate an advantage in total AMI optimization over conventional functions in terms of enabling the acquisition of constellations on the Pareto front of NMSE and noncoherent AMI. 
The constellation obtained using the proposed objective function significantly outperforms conventional Grassmann constellations in terms of channel estimation accuracy, thereby achieving superior performance in terms of the total AMI and allowing for improved spectral efficiency at middle SNRs compared with conventional RS.

\section*{Acknowledgment}
The authors would like to thank Dr. Yuto Hama from Ericsson Research, Japan, for providing valuable comments.

\footnotesize{
	\bibliographystyle{IEEEtranURLandMonthDiactivated}
	\bibliography{main,addition}
}

\end{document}